\documentclass[prb,twocolumn,aps,showpacs]{revtex4}
\usepackage{graphicx}%
\usepackage{dcolumn}
\usepackage{amsmath}
\usepackage{amssymb}
\usepackage{epsfig}

\begin{document}
\title{Anomalous dispersion of the collective modes  of an ultracold $^6Li-$ $^{40}K$ mixture in a square optical lattice }
\author{Zlatko  Koinov, Shanna Pahl, Rafael Mendoza }\affiliation{Department of Physics and Astronomy,
University of Texas at San Antonio, San Antonio, TX 78249, USA}
\email{Zlatko.Koinov@utsa.edu}
 \begin{abstract}

We report numerical calculations of the collective excitation spectrum and the speed of sound of the superfluid phase  of an atomic Fermi-Fermi mixture of population-imbalanced  Lithium-6 and Potassium-40 atoms in a square lattice. It is
predicted that in the exotic states of matter, known as the Fulde-Ferrell phase, an anomalous dispersion of the collective modes may be realized at  some values of polarization, interacting strength and temperature, i.e. the collective-mode dispersion initially bends upward before bending over as the quasimomentum increases.
\end{abstract}
 \maketitle
\section{Introduction}

The anomalous dispersion of the collective modes in a superfluid phase is a phenomenon which describes the upward deviation of the collective-mode spectrum from linearity. In this case, the conservation laws of energy and momentum allow processes which do not conserve the number of excitations, such as the three-particle process in which one particle decays into two with lower energy, or two interacting particles combining into one. The energy conservation condition requires that the dispersion relation first bends up as quasimomentum increases, before bending over. From an experimental point of view, the existence of anomalous dispersion can be confirmed by neutron scattering, which allows the direct determination of the collective-mode dispersion. However, neutron measurements
are difficult to be conducted because rather small scattering angles must
be employed. The anomalous dispersion is well established in superfluid $^4\textrm{He}$ for pressures $\leq 20$ bar.\cite{M1,M2,He1,He2} In this region one long-wavelength excitation (referred to as a superfluid phonon) can decay into another one by absorbing a second phonon (the Landau damping), or one long-wavelength phonon can decay  into two others (the Beliaev damping).

The subject
of anomalous dispersion of the collective modes  in a Bose gas in a periodic optical lattice potential at low temperatures has been explored in Refs. [\onlinecite{TG1,TG2}]. It was found that the spectrum of the collective modes of the Bose-Hubbard model exhibits an anomalous phonon dispersion under some critical on-site inter-atom interaction. This may cause Landau damping of collective modes in Bose condensates in a one-, two-, and three-dimensional periodic optical lattice potential.

 The question naturally arises as to  whether it is possible for the collective-mode dispersion of the Fermi-Hubbard model in the long-wavelength limit to convex concave up and thus exhibit an anomalous dispersion. Turning our attention to the theoretical description of the single-particle  and collective-mode excitations of superfluid alkali atom Fermi gases in
optical lattice potentials, we find that there have been  impressive theoretical achievements.\cite{SF0,SF1,SF1a,SF2,SF3,SF4,SF5,SF6,SF7,SF8,SF9,SF9a,SF9b,SF10,SF11,SF11a,SF12,SF13,SF14,SF15,SF15a,SF16,SF16a,SF17,SF17a,SF18,
SF19,SF19a,SF19b,SF19c,SF19dd,SF19d,SF19e,SF19g,SF20,SF21,SF22} Despite the fact that the  anomalous dispersion of the collective modes in Fermi condensates has not been predicted to exist in these papers, it was found that when the anomalous dispersion of interacting Fermi atoms is represented by the relation $\omega(Q)=cQ(1+\gamma Q^2-\delta Q^4)$, where $c$ is the speed of sound, and  $\gamma,\delta>0$, then the  scattering amplitude is formally equivalent  to the corresponding expression obtained in the case of three-phonon damping in superfluid $^4\textrm{He}$.\cite{MMT}

In what follows, we use the generalized random phase approximation (GRPA) to calculate the long-wavelength limit  of the collective  excitation spectrum, and the corresponding speed of sound, of  an  interacting Fermi mixture of Lithium-6 and Potassium-40 atoms in a two-dimensional optical lattice at finite
temperatures with the Fulde-Ferrell\cite{FF} (FF) order parameter. Our numerical calculations show that at some values of polarization, interacting strength and temperature the collective-mode dispersion $\omega(Q)$ initially bends upward before bending over as the quasimomentum $Q$ increases.
\section{Long-wavelength dispersion of the collective modes  of an ultracold $^6Li-$ $^{40}K$ mixture }

From a theoretical point of view, the simplest approach to the fermions in optical lattices is the tight-binding approximation,
 which requires a sufficiently deep lattice potential. In the tight-binding limit, two alkali atoms of opposite
pseudospins on the same site have  an interaction energy $U$, while
the probability to tunnel to a neighboring site is given by the
hopping parameters. The hopping parameters as well as the interaction
energy depend on the depth of the  lattice potential  and can be
 tuned  by varying the intensity of the laser beams. We  assume that the interacting
fermions are in a sufficiently deep periodic lattice potential
described by the Hubbard Hamiltonian. We  restrict the discussion to
the case of atoms confined to the lowest-energy band (single-band
Hubbard model), with two possible  states described by
pseudospins $\sigma$.  We consider different amounts of $^6Li$ and $^{40}K$ atoms in each  state ($\sigma=\uparrow=Li$, $\sigma=\downarrow=K$)
achieved by considering different chemical potentials $\mu_\uparrow$ and $\mu_\downarrow$. There are $M = M_\uparrowª+M_\downarrow$ atoms distributed along $N$ sites, and the corresponding filling
factors $f_{\uparrow,\downarrow}=M_{\uparrow,\downarrow}/N$ are smaller than unity. The Hubbard Hamiltonian is defined as follows:
\begin{equation}H=-\sum_{<i,j>,\sigma}J_{\sigma}\psi^\dag_{i,\sigma}\psi_{j,\sigma}
-U\sum_i \widehat{n}_{i,\uparrow}
\widehat{n}_{i,\downarrow}-\sum_{i,\sigma}\mu_\sigma\widehat{n}_{i,\sigma},\label{H}\end{equation}
where $J_{\sigma}$ is the single electron hopping integral,  and
 $\widehat{n}_{i,\sigma}=\psi^\dag_{i,\sigma}\psi_{i,\sigma}$ is the
density operator on site $i$. The Fermi operator
$\psi^\dag_{i,\sigma}$ ($\psi_{i,\sigma}$) creates (destroys) a
fermion on the lattice site $i$  with pseudospin projection
$\sigma$. The symbol $\sum_{<ij>}$ means sum over nearest-neighbor
sites of the two-dimensional lattice. The first term in (\ref{H}) is the usual kinetic
energy term in a tight-binding approximation. All numerical calculations will be performed  assuming that
 the hopping (tunneling) ratio $J_{Li}/J_K \approx 0.15$. In our notation  the strength of the on-site interaction $U>0$ is positive, but the
negative sign in front of the interaction corresponds to the Hubbard
model with an attractive interaction. In the presence of an
(effective) attractive interaction between the fermions, no matter
how weak it is, the alkali atoms form bound pairs, also called the
Cooper pairs. As a result, the system becomes unstable against the
formation of a new many-body  superfluid ground state. The
superfluid ground state comes from the U(1) symmetry breaking,  characterized by a nonzero order parameter, which in the
population-balanced case is assumed to be a constant in space
$\Delta_0$. Physically, it describes a superfluid state of  Cooper
pairs with zero momentum. A superfluid state of Cooper pairs with
nonzero momentum  occurs in the population-imbalanced case between a
fermion with momentum $\textbf{k} + \textbf{q}$ and spin $\uparrow$
and a fermion with momentum $-\textbf{k} + \textbf{q}$, and spin
$\downarrow$ . As a result, the pair momentum is $2\textbf{q}$. A
finite pairing momentum implies a position-dependent phase of the
order parameter, which in the  FF
case  varies as a single plane wave
$\Delta(\textbf{r})=\Delta_\textbf{q}\exp\left(2\imath
\textbf{q.r}\right)$, where $\Delta_\textbf{q}$ is a real quantity.
The order parameter  can also be a combination of two plane waves as
in the case of the Larkin-Ovchinnikov\cite{LO} (LO)
superfluid states.   In both cases, we are dealing with spontaneous
translational symmetry breaking and  an inhomogeneous superfluid
state. When continuous and global symmetries are spontaneously
broken the collective modes, known as the Nambu-Goldstone
 modes, appear.

 \begin{figure} \includegraphics[scale=0.75]{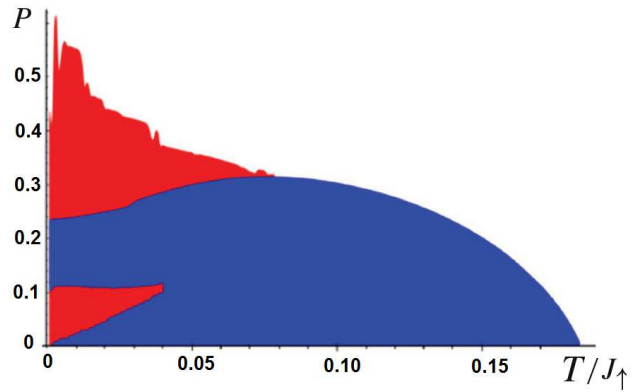}
    \caption{The phase diagrams of a $^6Li-$$^{40}K$ mixture in a square lattice [\onlinecite{SZ}]. The interaction strength is
$U = 2J_{Li}$.  The polarization is defined as $P = ( f_K - f_{Li} )/ f $,
where the total filling is $f = 0.5$ atoms/lattice site. Colors: Sarma states = blue (black), FF = red
(dark grey), and normal gas = white.   }\end{figure}

 The mean-field treatment of the FF and LO
phases in a variety of systems  shows that the FF and LO states
compete with a number of other states, such as the Sarma
($\textbf{q}=0$) states, and the superfluid-normal separation phase
(also known as the phase separation phase). It turns out that in
some regions of momentum space the FF (or LO) phase provides the
minimum of the mean-field expression of the Helmholtz free energy.  Phase diagrams for a $^6Li-$ $^{40}K$ mixture  at zero temperature were obtained in  Ref. [\onlinecite{SF11a}], but the calculations were limited to the emergence
of insulating phases during the evolution of superfluidity
from the BCS to the BEC regime, and the  competition between the FF and Sarma phases was
ignored.
The polarization versus temperature diagrams in  Fig. 1
show that there are three phases in the mass-imbalanced case : the Sarma phase, the FF phase, and
the normal phase in which the Helmholtz free energy is minimized for
gapless phase. The zero polarization line is the conventional
Bardeen-Cooper-Schrieffer state. Contrary to the phase diagram of
population-imbalanced $^6Li$ Fermi gas, where the phase separation
appears for low polarizations,  the existence of a
polarization window for the FF phase was found. This means that as soon as the
system is polarized it goes into the FF phase if the temperature is
low enough. This polarization window is larger for a majority of
$^{40}K$ atoms compared to the majority of $^6Li$ atoms.
Since the GRPA is a good approximation in a weak-coupling regime, we have chosen the on-site interaction to be $U/J_{Li}=2$. The mean-field number-, gap- and q-equations\cite{SZ} were solved at a temperature $T/J_{Li}=0.01$ for three polarizations: $P=(f_{K}-f_{Li})/(f_{K}+f_{Li})=0.1, 0.3$, and $P=0.4$, where the total filing factor is $f=f_{K}+f_{Li}=0.5$. The corresponding mean-field results for the Fulde-Ferrell wave vector $\textbf{q}=(\widetilde{q}\pi/a,0)$  ($a$ is the lattice constant), the two chemical potentials $\widetilde{\mu}_{Li,K}=\mu_{Li,K}/J_{Li}$ and the gap $\widetilde{\Delta}=\Delta/J_{Li}$ are as follows:
\begin{equation}\begin{split}&P=0.1, \widetilde{q}=0.049, \widetilde{\mu}_{Li}=2.091,\widetilde{\mu}_{K}=0.551,\widetilde{\Delta}=0.367,\\&
P=0.3, \widetilde{q}=0.122, \widetilde{\mu}_{Li}=1.894,\widetilde{\mu}_{K}=0.479,\widetilde{\Delta}=0.170,\\&
P=0.4, \widetilde{q}=0.153, \widetilde{\mu}_{Li}=1.683,\widetilde{\mu}_{K}=0.493,\widetilde{\Delta}=0.108.\label{MFP}\end{split}\end{equation}

\begin{figure} \includegraphics[scale=0.75]{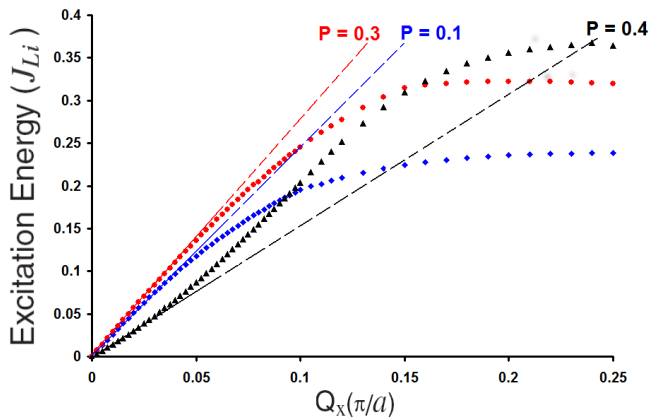}
    \caption{The collective-mode dispersion  $\omega(Q_x)$ in positive $Q_x$-direction calculated by the Bethe-Salpeter
    formalism for a 2D system with $U=2J_{Li}$
    and  $T=0.01J_{Li}$ and three different polarizations: $P=0.1$ - diamonds ( blue), $P=0.3$ - circles (red), and P=0.4 - triangles (black).  The red, blue and black straight lines define the slope of the curves in the long-wavelength limit and the corresponding speed of sound. The mean-field system parameters are given in the text. The anomalous dispersion appears for polarization P=0.4.   }\end{figure}

 \begin{figure} \includegraphics[scale=0.75]{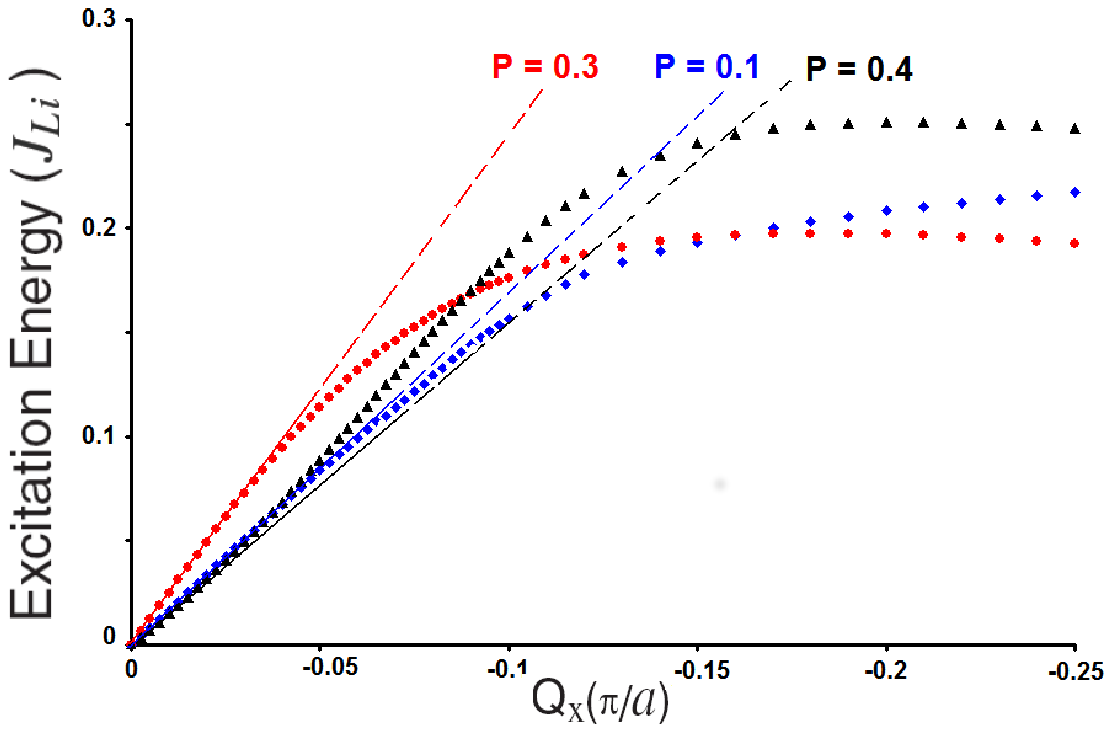}
    \caption{he collective-mode dispersion  $\omega(Q_x)$ in negative $Q_x$-direction calculated by the Bethe-Salpeter
    formalism for a 2D system with $U=2J_{Li}$
    and  $T=0.01J_{Li}$ and three different polarizations: $P=0.1$ - diamonds ( blue), $P=0.3$ - circles (red), and P=0.4 - triangles (black).  The red, blue and black straight lines define the slope of the curves  in the long-wavelength limit and the corresponding speed of sound.   The mean-field system parameters are given in the text. The anomalous dispersion appears for polarization P=0.4.   }\end{figure}
Generally speaking, the collective excitations of the Hamiltonian (\ref{H}) manifest themselves as poles of  the two-particle Green's function (or equivalently, as poles of the density and spin response functions).  Since the fermion self-energy does depend on the two-particle Green's function, the positions of  poles of the single-particle and two-particle Green's functions must  be obtained self-consistently. However, the GRPA is a widely accepted approximation in a weak-coupling regime. In this approximation the single-particle excitations and the corresponding single-particle
Green's functions are  calculated in the mean-field approximation, while the collective modes are
obtained by solving the Bethe-Salpeter (BS) equation. The kernel of the BS equation in the GRPA is obtained  by summing ladder and bubble diagrams. The mean-field decoupling of the single-particle and two-particle Green's functions leads to  expressions for the Green's functions that cannot be
evaluated exactly because the  interaction part
of the  Hamiltonian (\ref{H}) is quartic in the fermion fields. The simplest way to solve this problem is to
transform the quartic term
 into  quadratic form by making the
Hubbard-Stratonovich  transformation  for the fermion operators. In
contrast to the previous approaches, such that  after performing the
Hubbard-Stratonovich  transformation the fermion degrees of freedom
are integrated out; we decouple the quartic problem by introducing a
model system which consists of a multi-component boson field
interacting with  fermion fields.\cite{ZGK}

  The mean-field  single-particle Green's function, used in our numerical calculations, is a $4\times 4$ matrix, which takes into account  all possible   thermodynamic averages. The poles of the two-particle Green's function (the solutions of the  Bethe-Salpeter equation) are defined by the zeros of the corresponding $8\times 8$ secular determinant.\cite{ZGK}  It is worth mentioning, that it is possible  to reduce the single-particle Green's function to the $2\times 2$ one by neglecting some of the thermodynamic averages. As a result, the corresponding secular determinant reduces to a $4\times 4$ determinant.

We have calculated the collective-mode dispersion in $Q_x$-direction using the $4\times 4$ and the $8\times 8$ secular determinants at three different polarizations. The corresponding  mean-field system parameters are listed in  (\ref{MFP}). It turns out that the two  secular determinants provide almost the same collective-mode dispersion (the difference is about $2\%-7\%$ in the interval $-0.1\pi/a<Q_x<0.1\pi/a$, and less than $1\%$  out of this interval). The speed of sound, $c_{\pm}$, to the positive and negative directions of the $Q_x$  axis is defined by  $d\omega(Q_x)/dQ_x$ at $Q_x\rightarrow 0$.

In Fig. 2 and Fig. 3, we have presented the three collective-mode dispersions $\omega(Q_x)$  numerically calculated by using the $8\times 8$ secular determinant.   The straight lines define the slope of the dispersion curves and the corresponding speeds of sound:
 \begin{equation}\begin{split}&P=0.1,\quad c_{+}=0.614 J_{Li}a/\hbar,\quad c_{-}=0.534 J_{Li}a/\hbar\\&
P=0.3,\quad c_{+}=0.910J_{Li}a/\hbar,\quad c_{-}=0.793J_{Li}a/\hbar\\&
P=0.4,\quad c_{+}=0.466J_{Li}a/\hbar,\quad c_{-}=0.500J_{Li}a/\hbar.\nonumber\end{split}\end{equation}

 It can be seen, that at polarizations $P=0.1$ and $P=0.3$ the dispersion curves are under corresponding straight lines, and therefore we have normal dispersion: the phase velocity is less than the sound velocity. At polarization  $P=0.4$, the dispersion curve  initially bends upward before bending over as the quasimomentum increases. At some finite quasimomentum in positive and negative $Q_x$-directions, the phase velocity does exceed the corresponding sound velocity. At the momenta $Q_0\approx 0.23\pi/a$ and $Q_0\approx -0.15\pi/a$, the sound velocity equals the phase velocity and spontaneous decay is prohibited for momenta $Q_x>Q_0$. For small $Q_x$, the best-fit dispersion curve has no quadratic term in $Q_x$, and the dependence $\omega(Q_x)$ is similar to the corresponding fit in the  $^4\textrm{He}$ case:\cite{KU}
 $$\omega(Q_x)=uQ_x\left[1+\delta_1(Q_xa)^2-\delta_2(Q_xa)^3+\delta_3(Q_xa)^4+...\right].$$
In positive $Q-$direction we have: $u=0.478J_{Li}a/\hbar$, $\delta_1=9.78$, $\delta_2=24.37$, and $\delta_3=15.20$. In negative direction the parameters are: $u=0.496J_{Li}a/\hbar$, $\delta_1=10.44$, $\delta_2=36.03$, and $\delta_3=30.6$ It is worth mentioning, that this fit represents better the curve compare with the relation $\omega(Q)=cQ(1+\gamma Q^2-\delta Q^4)$ assumed in Ref. [\onlinecite{MMT}].

In conclusion, we have shown that collective excitation spectrum of the Fulde-Ferrell superfluid phase  of an atomic Fermi-Fermi mixture of population-imbalanced  Lithium-6 and Potassium-40 atoms in a square lattice may exhibit anomalous dispersion at  some values of polarization, interacting strength and temperature, and therefore, it is possible to have the damping of collective modes due to the three-particle process in which one particle decays into two with lower energy, or two interacting particles combine into one.

\end{document}